\documentclass[sigconf,screen, nonacm]{acmart}

\usepackage{listings}
\usepackage{microtype}

\usepackage{cleveref}
\usepackage{enumitem}

\lstdefinelanguage{Dafny}{
    keywords={method, function, class, constructor, trait, extends, returns, requires, ensures, modifies, invariant, if, else, true, false, while, for, match, case, var, const, int, nat, real, bool, bv4, char, string, array, set, multiset, seq, map, new, print},
    keywordstyle=\color{blue}, %
    identifierstyle=\color{black},
    comment=[l]{//},
    commentstyle=\color{gray}\ttfamily,
    stringstyle=\color{red}\ttfamily,
    sensitive=true
}

\newcommand{\tool}{\textsf{DSpec2Test}}

\AtBeginDocument{%
  }

\begin{document}
\balance

\title{DSpec2Test: Specification-Driven Test Generation in Dafny}

\author{Sofia Vieira Pinto}
\orcid{0009-0001-0055-3129}
\affiliation{%
  \institution{INESC TEC, Faculdade de Engenharia, Universidade do Porto}
  \city{Porto}
  \country{Portugal}
}
\email{sofia.p.pt@gmail.com}

\author{Álvaro F. Silva}
\orcid{0009-0005-2941-9942}
\affiliation{%
  \institution{INESC TEC, Faculdade de Engenharia, Universidade do Porto}
  \city{Porto}
  \country{Portugal}
}
\email{amfpsilva@hotmail.com}

\author{João Pascoal Faria}
\orcid{0000-0003-3825-3954}
\affiliation{%
  \institution{INESC TEC, Faculdade de Engenharia, Universidade do Porto}
  \city{Porto}
  \country{Portugal}
}
\email{jpf@fe.up.pt}

\author{Alexandra Mendes}
\authornote{Author order follows a senior-author-last convention.}
\correspondingauthor
\orcid{0000-0001-8060-5920}
\affiliation{%
  \institution{INESC TEC, Faculdade de Engenharia, Universidade do Porto}
  \city{Porto}
  \country{Portugal}
}
\email{afmendes@fe.up.pt}

\renewcommand{\shortauthors}{Sofia Vieira Pinto, Álvaro F. Silva, João Pascoal Faria, and Alexandra Mendes}

\begin{abstract}
Verification-aware languages, such as Dafny, integrate logical constructs into code and enable automatic verification of program correctness.
However, tests remain helpful in scenarios that verification
alone does not address (e.g., to support test-driven development). Existing Dafny test generation tools are
implementation-based, limiting their applicability in this context.

We present \tool, a specification-driven test generation tool for Dafny that automatically derives tests from formal specifications, without considering implementation details. Our tool extends Dafny's \texttt{generate-tests} command with a new black-box mode based on Disjunctive Normal Form (DNF) equivalence class partitioning and optional Boundary Value Analysis (BVA). \tool~relies on the Z3 SMT solver to synthesize inputs and expected outputs that meet the specification-derived constraints.

We evaluate \tool~on programs from DafnyBench mutated using MutDafny and compare it against Dafny's existing implementation-driven \emph{Block} mode. \tool~achieves a 93.9\% mutation kill rate on a dataset of 131 mutants, outperforming \emph{Block}'s 82.4\%, and uniquely killing 17 mutants. These results suggest that specification-driven testing is an effective and complementary approach for testing Dafny programs.

\noindent
\textbf{Demo:} \href{https://youtu.be/mK1EeJfinRQ}{\ttfamily \small  https://youtu.be/mK1EeJfinRQ}\\
\textbf{Code:} \href{https://github.com/VeriFixer/DSpec2Test}{\ttfamily \small https://github.com/VeriFixer/DSpec2Test}\\
\textbf{Prebuilt docker image:}
\href{https://doi.org/10.5281/zenodo.21191158}{\ttfamily \small  https://doi.org/10.5281/zenodo.21191158}
\end{abstract}




\maketitle

\section{Introduction}

Verification-aware languages, such as Dafny~\cite{dafny}, integrate preconditions (\emph{requires}), postconditions (\emph{ensures}), and invariants (\emph{invariant}) directly into the code, enabling automated detection of specification violations during development.
However, when verification fails, error messages are often not sufficiently informative to identify the underlying fault~\cite{oliveira2025challenges},
costing developers significant time and effort. Here, tests can serve as a complementary mechanism by providing concrete executions that help locate and understand the fault.

Within Dafny, test generation from implementation code is already supported by DTest~\cite{fedchin2023toolkit}, accessible via the \texttt{generate-tests} command.  DTest targets code coverage to provide additional assurance that compilation of Dafny programs to target languages (e.g., C\# or Java) preserves the properties verified in Dafny, attempting to mitigate the gap left by Dafny's unverified compilation
pipeline. It is part of a broader toolkit that also includes DUnit (unit testing) and DMock (mock testing).

While DTest is useful for validating compiled implementations, there are scenarios in which generating tests directly from specifications is more suitable. Implementation-based tests are constrained by the structure and behaviour of the existing code, whereas tests based on the specification focus on the intended semantics of the program. This distinction is especially relevant during early stages of development, where specifications may already exist but implementations are still incomplete or evolving. In such settings, specification-driven tests support a test-driven development (TDD) style, where tests are derived from the specification first and guide the subsequent implementation. This can help to pinpoint errors as soon as implementations deviate from the intended behavior.

We present \tool, a tool for specification-driven test generation in Dafny. This tool is provided through an additional mode (\emph{Spec}) for Dafny’s \texttt{generate-tests} command. It analyses the preconditions and postconditions of the methods and functions to create Disjunctive Normal Form (DNF) clauses, producing a comprehensive pool of tests.
To the best of our knowledge, there are no other Dafny tools that allow the automatic creation of tests prior to the program's implementation being written.

Our main \textbf{contributions} are:
\begin{enumerate}[leftmargin=*]
\item[\small{$\bigstar$}] A specification-to-test pipeline (instantiated for Dafny) that combines (i)~an \emph{automatic} \emph{Safe DNF} decomposition of arbitrary pre/postconditions, preserving
short-circuit guard semantics, (ii)~boundary value analysis on
the partition-derived constraints, and (iii)~an encoding of
each test goal as an \texttt{assume}/\texttt{assert false}
stub that delegates the search for concrete inputs and
expected outputs to Z3.
   \item[\small{$\bigstar$}] \tool, the first tool for \emph{specification-based conformance testing} in Dafny. %
   It derives tests from
\texttt{requires}/\texttt{ensures} clauses and can support
test-driven development, specification validation, and
fault localisation.
   \item[\small{$\bigstar$}] Evidence that specification-based and implementation-based testing reach complementary fault classes gathered through an empirical evaluation on 131 %
   mutants from DafnyBench, in which \tool~kills 93.9\,\%, outperforming DTest (82.4\,\%) and uniquely killing 17 mutants that DTest misses.
\end{enumerate}

\section{Formal Specification-Driven Test Generation}\label{sec:methodology}
Formal specification-driven test generation is an approach that constructs test cases directly from formal specifications rather than relying on existing implementation code.
\tool~follows this approach by actively exercising all preconditions (\emph{requires}) and postconditions (\emph{ensures}), and by automatically applying the well-established black-box testing strategies of equivalence class partitioning (ECP) and boundary value analysis (BVA) ~\cite{myers2004art}. %
\tool~relies on an SMT-solver (Z3) to find concrete test data that satisfy the target test conditions produced by ECP and BVA.

\subsection{DNF-Based Equivalence Class Partitioning}

ECP consists of dividing input data into partitions, called
equivalence classes, where all values in a class are expected
to behave in the same way~\cite{myers2004art}. We automate ECP
by analysing the method's contract clauses and converting them
into Disjunctive Normal Form (DNF). Alternative behaviours
(depending on different input classes) are typically encoded
in the method's postconditions, which is why our starting
point is the conjunction of all preconditions (\texttt{requires})
and postconditions (\texttt{ensures}). Each DNF branch then
constrains both inputs and outputs, allowing us to generate
concrete values for the test inputs and expected outputs (when
uniquely determined by the contract).

\begin{lstlisting}[language=Dafny, caption={Example Dafny contract: \texttt{ConditionalGet}.}, label=lst:dafny-program, basicstyle=\fontsize{8}{9}\ttfamily]
method ConditionalGet(s: seq<int>, x: int) returns (r: int)
  requires |s| > 0
  ensures x < 0 ==> r == -1              // Q1
  ensures x >= |s| ==> r == 1            // Q2
  ensures 0 <= x < |s| ==> r == s[x]   // Q3
\end{lstlisting}

Conversion to DNF must be done carefully: a naive expansion of
an implication $A \Rightarrow B$ as $\neg A \lor B$ produces a
branch in which $B$ is evaluated without the guard $A$, which
may cause runtime crashes. For example, naively expanding $Q_3$
in \Cref{lst:dafny-program} yields a branch where $r = s[x]$ is
evaluated without the guard $0 \leq x < |s|$, raising an
out-of-bounds error. To avoid this, \tool's default
\emph{Safe DNF} mode applies the short-circuit-safe
decomposition rules of~\Cref{tab:dnfrules}, which produce
mutually exclusive, guarded branches.

\begin{table}[h]
\centering
\caption{Safe DNF decomposition rules.}
\label{tab:dnfrules}
\begin{tabular}{ll}
\hline
\textbf{Operator} & \textbf{Branches (mutually exclusive)} \\
\hline
$A \Rightarrow B$            & $\{\,\neg A,\; A \land B\,\}$ \\
$A \lor B$                   & $\{\,A,\; \neg A \land B\,\}$ \\
$A \Leftrightarrow B$        & $\{\,A \land B,\; \neg A \land \neg B\,\}$ \\
\texttt{if A then B else C}  & $\{\,A \land B,\; \neg A \land C\,\}$ \\
\hline
\end{tabular}
\end{table}

\noindent
Applied to \texttt{ConditionalGet}, each implication splits into two cases:
\begin{itemize}[leftmargin=*]
\item $Q_1$: $\{\, x \geq 0,\; x < 0 \land r = -1 \,\}$
\item $Q_2$: $\{\, x < |s|,\; x \geq |s| \land r = 1 \,\}$
\item $Q_3$: $\{\, \neg(0 \leq x < |s|),\; 0 \leq x < |s| \land r = s[x] \,\}$
\end{itemize}
The cross product, conjoined with the precondition
$|s| > 0$, yields eight candidate equivalence classes. Five are
infeasible (Z3 returns UNSAT) due to mutually exclusive guards. The three feasible classes, after simplifying the redundant guard
negations, are:
\begin{itemize}[leftmargin=*]
\item $C_1$: $|s| > 0 \;\land\; x < 0 \;\land\; r = -1$        \quad // below range
\item $C_2$: $|s| > 0 \;\land\; x \geq |s| \;\land\; r = 1$    \quad // above range
\item\label{item:C3} $C_3$: $|s| > 0 \;\land\; 0 \leq x < |s| \;\land\; r = s[x]$  \quad // in range
\end{itemize}

If exhaustive combination testing is required and can be safely applied, \tool~also
supports \emph{Full DNF} (\texttt{--fdnf} flag), which drops
short-circuit safety and forces every $n$-ary disjunctive
operator to produce all $2^n - 1$ non-empty subsets of branch
satisfaction.
For instance, $A \lor B$ expands to $\{A \land B,\, A \land \neg B,\, \neg A \land B\}$.

\subsection{Boundary Value Analysis}

While equivalence class partitioning provides broad coverage,
defects (such as off-by-one errors) often hide at the edges of these partitions. Therefore, when the \texttt{--bva} flag is provided, the tool complements the DNF phases with Boundary Value Analysis (BVA).

For every partition, the tool generates multiple test
scenarios. The first test leaves all variables free, relying
only on the base DNF constraints.
Subsequent tests pin exactly one variable to a boundary value at a time, leaving all others free. This prevents conflicting constraints while exploring extreme edge cases.

The pinned boundary values are determined by each partition's
constraints. If the partition restricts a variable to an interval, the tool pins it to its exact valid edges.
When a variable is unbounded, the tool uses default extremes (-100 and 100) for numeric variables. Sequences, sets, and maps are tested for lengths $0$, $1$, and $>1$.

For instance, applied to class $C_3$ %
of \texttt{ConditionalGet}, BVA
emits four additional scenarios, with the following constraints:
$x = 0$ (minimum), $x = |s|{-}1$ (maximum), $|s| = 1$ (minimum), and $|s| > 1$ (above minimum).

\subsection{Concrete Test Data Generation}

Like DTest, \tool~relies on Z3 %
to find concrete test data that satisfy each
DNF branch and BVA constraint. The relevant constraints are encoded as \texttt{assume} statements in the method's body, followed by an \texttt{assert false} that forces the verifier to emit
a counterexample, providing concrete test inputs and expected outputs. For class $C_3$ of \texttt{ConditionalGet}, the instrumented body is:

\begin{lstlisting}[language=Dafny, basicstyle=\fontsize{8}{9}\selectfont\ttfamily, caption={Instrumented body for class $C_3$.}, label=lst:probe-c3]
method ConditionalGet(s: seq<int>, x: int) returns (r: int)
  // contract unchanged
{
  assume |s| > 0 && 0 <= x < |s| && r == s[x];
  assert false;
}
\end{lstlisting}

\noindent
A counterexample such as \texttt{s=[3]}, \texttt{x=0}, \texttt{r=3} is materialised as a runnable Dafny test that sets up the input parameters, calls the method under test, and checks its outputs against concrete expected values produced by Z3 (when \texttt{--simplify} is set, and the postconditions uniquely constrain the outputs) or the original postconditions instantiated for the known inputs.

\begin{lstlisting}[language=Dafny, basicstyle=\fontsize{8}{9}\selectfont\ttfamily, caption={Generated test with \texttt{--simplify} flag for class $C_3$.}, label=lst:test-c3]
method {:test} Test_ConditionalGet_C3() {
  var r := ConditionalGet([3], 0);
  expect r == 3;
}
\end{lstlisting}

\subsection{Implementation Pipeline}

\begin{figure}[!htbp]
    \includegraphics[width=0.47\textwidth]{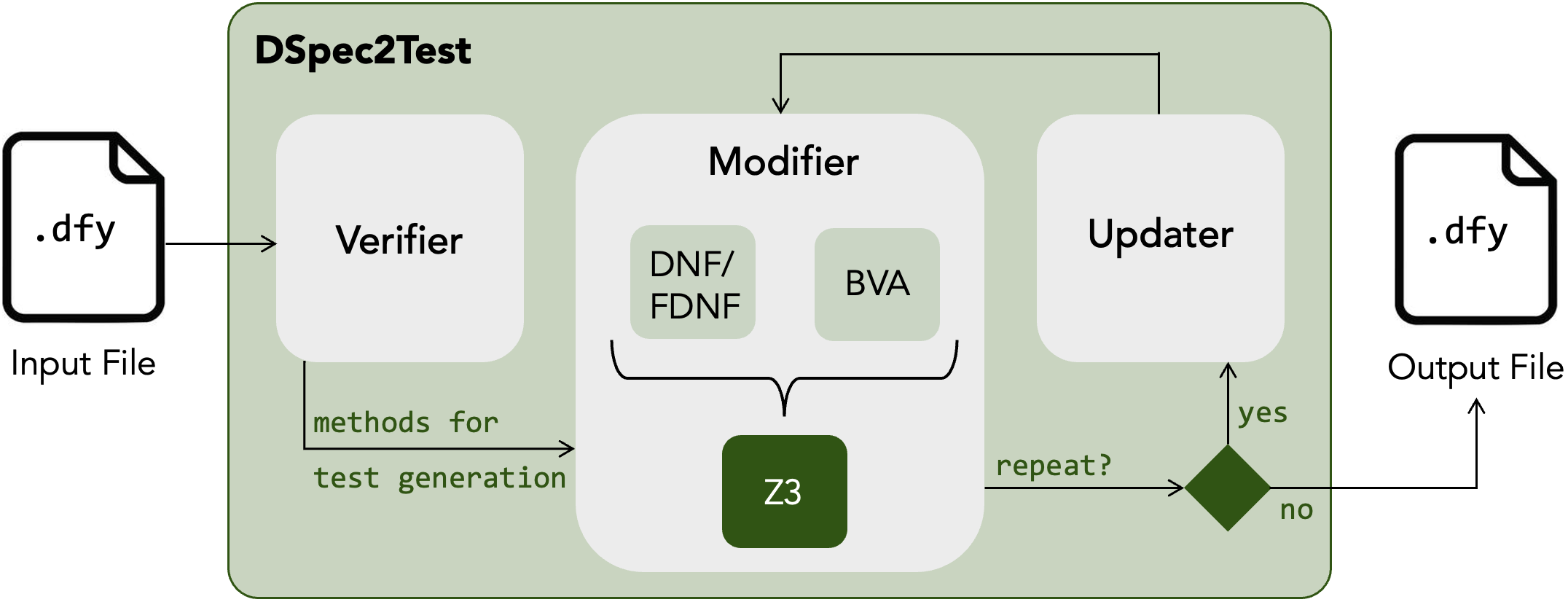}
    \caption{\tool~pipeline. An input file is processed by the \emph{Verifier}, where methods that are marked with \texttt{\{:testEntry\}} or fail verification are passed to the \emph{Modifier} for DNF generation and optional BVA. The \emph{Updater} iteratively refines the process by generating additional tests while ensuring novelty through added \texttt{assume} clauses.}
    \label{fig:diagram}
\end{figure}

\Cref{fig:diagram} summarises how the components described
above are integrated into Dafny's \texttt{generate-tests}
command, in three stages.

The \emph{Verifier} flags for test generation every method that
fails verification or is annotated with \texttt{\{:testEntry\}}.
The implementations of these methods are then discarded, since
they are not used for specification-based test generation.

The \emph{Modifier} instruments each flagged method with the
DNF classes and, when \texttt{--bva} is selected, BVA boundary
pins derived in the previous subsections, as illustrated in
\Cref{lst:probe-c3}.

The \emph{Updater} drives iteration when \texttt{--repeat} is higher than one and \texttt{--bva} is not selected. After each
generated test, it appends a fresh \texttt{assume} clause that
excludes the previous input, so that the next solver call
returns a different one. For instance, if \texttt{x} was
assigned to \texttt{0}, an \texttt{assume x != 0;} statement
is added before the next iteration.

\section{Empirical Evaluation}

\subsection{Methodology}
We compare two strategies of Dafny's \texttt{generate-tests}:
\textbf{\emph{Block}}, the existing white-box mode provided by
DTest, which targets path/branch coverage of the
implementation; and
\textbf{\emph{Spec\_bva}}, the black-box mode introduced by
\tool~(\emph{Spec}) with the \texttt{--bva} flag, which targets equivalence class and boundary value
coverage of the specification.

We evaluate fault-detection capability on a corpus of 20 programs sampled from DafnyBench~\cite{loughridge2024dafnybench}, selected to contain methods with formal specifications (pre/postconditions, loop invariants) amenable to automated test generation. Currently, neither of the tools (DTest and \tool) supports mutable state (e.g., arrays), class fields, and non-determinism (e.g., \texttt{:|} operator), so these constraints were also considered during the sampling process. Each program was annotated with \{:testEntry\} so tests could be generated from the original (unmutated) code.

We applied mutation testing, a well-established technique for evaluating test adequacy~\cite{papadakis2019mutation, petrovic2018state}, using MutDafny~\cite{amaral2025mutdafny}, a mutation tool that injects syntactic faults (relational/arithmetic operator replacement, statement deletion, conditional negation, etc.) into Dafny implementations. For each program, up to 10 mutants were generated, retaining only those that fail verification, %
yielding 200 mutants in total. A mutation that fails verification may preserve observable behaviour\,---\,it may simply break verification constructs. We mark such cases as \emph{equivalent} mutants.

The evaluation pipeline, using Dafny 4.11.0 and Z3 4.12.1, (1)~generates tests from the original program, (2)~runs a safety check ensuring that the tests pass on the original (unmutated) code, and (3)~runs each suite against every mutant to determine kill/survive/timeout status.

\subsection{Results and Discussion}

We base the comparison on the 17/20 programs supported by both strategies (three were not supported, crashing during test creation for at least one of them due to either exponential complexity or tests not meeting the preconditions), totaling 170 mutants.

\noindent
\textbf{\emph{Analysis of surviving mutants.}} We manually inspected the mutants that survived both strategies to determine if they could be realistically killed, and found 22 \emph{non-compiling} mutants and 17 \emph{equivalent} mutants (observationally equivalent to the original during execution), leaving \emph{131 realistically-killable mutants}. %

\noindent
\textbf{\emph{Analysis of timeouts.}} A small number of mutants timed out at the 300\,sec kill-check budget, which is vastly longer than a standard Dafny test execution. A manual inspection confirmed that these were all caused by infinite loops introduced by the mutations, exposed when exercised by the generated test inputs. We therefore credit timeouts as kills.

\noindent
\textbf{\emph{Kill rates.}} Table~\ref{tab:results} reports per-strategy kill rates and the overlap between the two tools on the 131 that can realistically be killed.
\emph{Spec\_bva} kills 123 (93.9\%), \emph{Block} kills 108 (82.4\%); the union covers 125 (95.4\%), with \emph{Spec\_bva} uniquely killing 17, \emph{Block} uniquely killing 2, and only 6 mutants killed by neither tool.

\begin{table}[h]
\centering
\caption{Mutation kill rates and kill overlap on 131 realistically-killable mutants. Timeouts (TO) are runtime infinite loops credited as kills. Killed = FAIL + TO.}
\label{tab:results}
\begin{tabular}{lccccc}
\hline
& \textbf{FAIL} & \textbf{TO} & \textbf{Killed} & \textbf{Surv.} & \textbf{Kill\%} \\
\hline
\emph{Block} (WB)        &  99 &  9 & 108 & 23 & 82.4\% \\
\emph{Spec\_bva} (BB)    & 106 & 17 & 123 &  8 & \textbf{93.9\%} \\
Union             & --- & --- & 125 &  6 & 95.4\% \\
Killed by \emph{Block} only     & --- & --- &   2 & --- &  1.5\% \\
Killed by \emph{Spec\_bva} only & --- & --- &  17 & --- & 13.0\% \\
\hline
\end{tabular}
\end{table}

\noindent
\textbf{\emph{Test volume vs.\ kill yield.}}
\emph{Spec\_bva} generates 6.6 tests per file on average (860 across the 131 files), versus ${\sim}1.6$ for \emph{Block}. However, this volume is heavily redundant: extracting the smallest failing test index per mutant from the runner output, we find 52\,\% of explicit-failure kills come from the first test, and 73\,\% from the first three.

\noindent
\textbf{\emph{Complementarity.}}
Of the 19 mutants killed by exactly one tool out of 131, 17 (89\,\%) are unique to \emph{Spec\_bva} and 2 are unique to \emph{Block}, indicating that ECP and BVA on the specification reaches faults that path coverage on the implementation does not, and vice-versa, highlighting their complementarity.

\noindent
\textbf{\emph{Cost.}} \emph{Spec\_bva} is approximately 1.6$\times$ slower in wall-clock time (7889\,s vs.\ 4867\,s), reflecting the larger generated suite.

\section{Related Work}

In addition to DTest~\cite{fedchin2023toolkit},
Delfy~\cite{christakis2016integrated} also generates tests for
Dafny, using dynamic symbolic execution with Z3 to diagnose
verification failures. Concolic tools such as
PathCrawler~\cite{williams2005pathcrawler} (for C, within
Frama-C~\cite{kirchner2015frama}) similarly generate inputs
from executable code for structural coverage. Our approach
differs from these as it is fully spec-driven and requires no
implementation.

AutoTest~\cite{meyer2009AutoTest} is a black-box testing framework for Eiffel that combines random sampling with existing contracts to generate tests for object-oriented code. We instead use SMT solver calls for a more directed
exploration of the input space.

Other spec-driven test generators apply related techniques to
formal contracts. Gil et al.~\cite{gil2024automatic} target
SPARK/Ada (also a verification-aware language) with ECP guided
by user-written \texttt{Contract\_Cases} and Z3py called
externally, but, unlike our approach, do not automatically derive DNF-based
equivalence classes from arbitrary pre/postconditions or reuse
the host verifier. Peña et al.~\cite{pena2023smt} use
Z3 to derive test inputs from preconditions in a custom
specification language, but do not partition the whole contract
via DNF for directed, human-readable tests. Rebello
de Andrade et al.~\cite{rebello2012specification} apply Full
DNF to axiomatic specifications of Java generics --- a
precursor to our Safe DNF for method contracts --- and use Alloy Analyzer (a SAT solver) to find satisfying models for each minterm, but do not
target a verification-aware language or apply boundary value
analysis.

Property-based testing tools, such as
QuickCheck~\cite{claessen2000quickcheck} and Hypothesis~\cite{maciver2019hypothesis}, also generate tests from
properties, but these are executable, partial, and
input-centric, not full correctness specifications.

\section{Future Work}

We plan to refine DNF clauses with additional constraints so
that every postcondition literal genuinely constrains the
admissible outputs for the input Z3 picks. Consider finding
the last occurrence of $x$ in a sorted sequence $a$ known to
contain it, as shown in the example in \Cref{lst:lastidx}.

\begin{lstlisting}[language=Dafny, caption={Last occurrence of $x$ in a sorted sequence.}, label=lst:lastidx]
method LastIdx(a: seq<int>, x: int) returns (i: int)
 requires sorted(a) && x in a
 ensures 0 <= i < |a|         // Q1 (range guard)
 ensures a[i] == x             // Q2
 ensures x !in a[i+1..]        // Q3
\end{lstlisting}
Z3 picks the minimal model \texttt{a=[x]}, \texttt{i=0}, on
which $Q_2$ and $Q_3$ hold vacuously; a buggy implementation
returning \texttt{a[0]} passes. Exercising both non-vacuously
requires multiple copies of $x$ plus a distinct element
(e.g., \texttt{a=[5,5,6]}, \texttt{x=5}). A \emph{relevance
check} forces this: for each output-referring $Q_k$ (excluding
range guards), search for an alternate $i_k\!\neq i$
satisfying the other literals but violating $Q_k$.
Equivalently, satisfy the extended contract, as shown in \Cref{lst:lastidx-rel}.
\begin{lstlisting}[language=Dafny, caption={Extended contract for relevance checking.}, label=lst:lastidx-rel]
method LastIdx'(a: seq<int>, x: int)
               returns(i: int, ghost i2:int, ghost i3:int)
 requires sorted(a) && x in a
 ensures 0 <= i < |a| && a[i] == x && x !in a[i+1..]
 ensures i2!=i && 0<=i2<|a| && a[i2]!=x && x !in a[i2+1..] //Q2 rel.
 ensures i3!=i && 0<=i3<|a| && a[i3]==x && x in a[i3+1..] //Q3 rel.
\end{lstlisting}
\tool~also lacks \emph{function inlining} in specifications and (like DTest) \emph{support for mutable state} (arrays
and class fields). Both are planned to broaden applicability.

We also plan to expand the evaluation to a broader sample of programs.

\section{Conclusion}

We presented \tool, a specification-driven test generation tool for Dafny that derives runnable tests from the \texttt{requires} and \texttt{ensures} clauses of a method, without examining the implementation. While specification-based testing is well-established for other contract languages (e.g.,~AutoTest for Eiffel, JMLUnit for Java), \tool~is, to the best of our knowledge, the first Dafny tool that supports test generation prior to (or independently of) the implementation\,---\,a capability complementary to the existing implementation-driven DTest. As far as we are aware, it is also the first tool that performs automatic DNF partitioning of method contracts. %

Our evaluation shows that DTest and \tool~are complementary, with \tool~uniquely killing 17 mutants that \emph{Block} misses out of 131, indicating that specification- and implementation-based testing reach different fault classes. \tool~is therefore
useful both as a stand-alone tool for early-stage and TDD-style
workflows and as a complement to existing white-box tooling such as DTest, useful for implementation coverage. However, for maximum fault localisation and rigor, both strategies should be employed.

\section*{Data Availability Statement}

The data associated with this work are publicly available through a Zenodo repository~\cite{vieira_pinto_2026_21191158}. The dataset can be accessed using the following DOI: \url{https://doi.org/10.5281/zenodo.21191158}

\begin{acks}

Sofia Vieira Pinto, João Pascoal Faria, and Alexandra Mendes were funded by National Funds through the FCT - Fundação para a Ciência e a Tecnologia, I.P. (Portuguese Foundation for Science
and Technology) within the project VeriFixer, with reference 2023.15557.PEX (DOI: 10.54499/2023.15557.PEX).

Álvaro F. Silva was co-financed by national funds through FCT – Fundação para a Ciência e a Tecnologia, I.P., under the support UID/50014/2025 (https://doi.org/10.54499/UID/50014/2025), by National Funds through the FCT - Fundação para a Ciência e a Tecnologia, I.P. (Portuguese Foundation for Science
and Technology) within the project VeriFixer, with reference 2023.15557.PEX (DOI: 10.54499/2023.15557.PEX), and Fundação para a Ciência e a Tecnologia (Portuguese Foundation for Science and Technology) through the Carnegie Mellon Portugal Program under the fellowship reference PRT/BD/155045/2024.

\end{acks}

\bibliographystyle{ACM-Reference-Format}
\bibliography{references.bib}

\end{document}